\documentclass[11pt]{myclass}
 \usepackage{euscript,xcolor,citesort}
 
 \newcommand{\f}[1]{\ensuremath{\EuScript{#1}}}

 \title{A Gentle\footnote{This is a ``gentle'' introduction, which means that where necessary, it will sacrifice mathematical rigor for ease of exposition. The cited references provide a more rigorous treatment of the material.}~ 
 Introduction to the Kernel Distance}
 \author{Jeff M. Phillips \and Suresh Venkatasubramanian}

\begin{document}

\maketitle

\begin{abstract}
This document reviews the definition of the kernel distance, providing a gentle introduction tailored to a reader with background in theoretical computer science, but limited exposure to technology more common to machine learning, functional analysis and geometric measure theory.  
The key aspect of the kernel distance developed here is its interpretation as an $L_2$ distance between probability measures  or various shapes (e.g. point sets, curves, surfaces) embedded in a vector space (specifically an RKHS).  This structure enables several elegant and efficient solutions to data analysis problems.  
We conclude with a glimpse into the mathematical underpinnings of this measure, highlighting its recent independent evolution in two separate fields.  
\end{abstract}

\section{Definitions}

Let $K : \reals^d \times \reals^d \rightarrow \reals$ be a \emph{similarity function} with the property that for any $x, K(x,x) = 1$, and as the distance between $x$ and $y$ increases, $K(x,y)$ decreases. A simple example of a kernel is the \emph{Gaussian kernel} $K(x,y) = \exp(-\frac{\|x - y\|^2}{\sigma^2})$. 

\begin{defn}[Kernel Distance~\cite{Suq95,HB05,glaunesthesis,Vaillant2005,DBLP:conf/miccai/DurrlemanPTA08,DBLP:conf/miccai/DurrlemanPTA07,GlaunesJoshi:MFCA:06}]
  Let $P$ and $Q$ be sets of points in $\reals^d$. The \emph{kernel distance} between $P$ and $Q$ is 
\begin{equation}
\label{eq:current}
  D^2_K(P, Q) \triangleq \sum_{p \in P} \sum_{p' \in P} K(p, p')  + \sum_{q \in Q} \sum_{q' \in Q} K(q,q') - 2 \sum_{p \in P} \sum_{q \in Q} K(p,q).
\end{equation} 
\end{defn}

Note that the kernel distance is defined in terms of its square. We will continue this formulation so as to avoid the repeated use of square roots. While $D_K(P,Q)$ satisfies symmetry and one side of the identity of indiscernables, it might not in general be a metric, or even a pseudometric. In Section~\ref{sec:rkhs} we will discuss specific conditions on the similarity function $K$ for which $D_K$ is a metric; in short, $K$ must be a \emph{positive definite kernel}.

\section{From Similarities to Distances}

The construction of the kernel distance involves a transformation from similarities to distances. This transformation takes the following general form. Given two ``objects'' $A$ and $B$, and a measure of similarity between them given by $K(A,B)$, then the induced distance between $A$ and $B$ can be defined as the difference between the self-similarities $K(A, A) + K(B,B)$ and the cross-similarity $K(A,B)$:
\[ d(A, B) = K(A, A) + K(B,B) - 2K(A,B). \]
(Note that the multiplicative factor $2$ is needed to ensure that $d(A,A) = 0$.)  This construction can also be motivated set-theoretically: we can express the cardinality of the symmetric difference $S \Delta S'$ between two sets $S, S'$ (a measure of the distance between them) as the expression $|S \Delta S'| = |S| + |S'| - 2 |S \cap S'|$, where now the cardinality of the intersection plays the role of the similarity function $K$.

\subsection{Point-wise Similarities}
Consider two points $p, q$. The kernel distance $D^2_K(\{p\}, \{q\}) = K(p,p) + K(q,q) - 2K(p,q) = 2(1 - K(p,q))$ (when $K(p,p)=1$). Here, the expression $1-K(p,q)$ acts like a (squared) distance between the two points. Alternately, we can view the kernel distance as given by the sum of the \emph{self-similarities} of $p$ and $q$, from which we subtract the \emph{cross-similarity} $K(p,q)$. Note that 

\begin{itemize}
  \item if the two points are identical, this expression is zero. i.e $x = y \Rightarrow D_K(x,y) = 0$.
  \item $D_K(p,q) = D_K(q,p).$
  \item As the points become more distant, the self-similarities remain the same while the cross-similarity decreases, which increases $D_K$. 
\end{itemize}

Formulating the kernel distance in terms of self- and cross-similarities allows us to recover the general form of the definition. Let the \emph{cross-similarity} of point sets $P$ and $Q$ be given by $\kappa(P,Q) = \sum_{p \in P}\sum_{q \in Q} K(p,q)$. Then we can write 
\[ D^2_K(P,Q) = \kappa(P,P) + \kappa(Q,Q) - 2\kappa(P,Q). \]

\subsection{Comparing data with uncertainty}

Another way of interpreting the kernel distance comes from modeling data with uncertainty. If we have two sets of points $P$ and $Q$, a naive way of estimating their difference would be to compute the symmetric difference $|P \Delta Q| = |P| + |Q| - 2|P \cap Q|$. Of course, in all but a few cases, this number would be meaningless, since $|P \cap Q|$ would equal zero unless points exactly coincided. We can replace the sharp similarity function $k(p,q) = \mathbf{1}_{p=q}$ by the smoother similarity function $K(p,q)$, where $K(p, \cdot)$ represents uncertainty in the location of $p$ (for example, as a Gaussian). In this case, the above symmetric difference $|P \Delta Q|$ then turns into $D^2_K(p, q)$.

\section{Generalizations}
\label{sec:variations}
One of the features of the kernel distance is that it can be generalized beyond point sets to distributions in space and even to higher dimensional geometric structures like curves and surfaces, where it is often called the \emph{current distance}~\cite{Vaillant2005}. In this section we merely define the appropriate generalizations, giving some simple intuition where appropriate. A deeper treatment of the mathematics underpinning these constructions will be presented in Section~\ref{sec:motivation}. 

\subsection{Measures and Distributions}
We can think of a point set as a set of $\delta$-functions at each location, which gives us a representation of the point set as a measure over $\reals^d$. We can then weight these functions, which allows us to compare \emph{weighted} point sets. Let $\f{P} = (P, w)$ be a point set with an associated weight function $w : P \rightarrow \reals$, and let $\f{Q} = (Q, w')$ be another such weighted point set. We can define the cross-similarity $\kappa(\f{P}, \f{Q}) = \sum_{p \in P} \sum_{q \in Q} w(p)K(p,q) w'(q)$ and then as before define 
\[ D^2_K(\f{P}, \f{Q}) = \kappa(\f{P}, \f{P}) + \kappa(\f{Q}, \f{Q}) - 2 \kappa(\f{P}, \f{Q}).\]
In general, given two distributions over $\reals^d$, we can define the kernel distance between them merely by replacing the sum in the cross similarity by an integration. Specifically, if we are given two distributions $\mu, \nu$, we define the cross-similarity as 
\[ \kappa(\mu, \nu)  = \int \int K(p,q) d\mu(p) d\nu(q) \]
after which $D_K$ can be defined in the usual manner.

\subsection{Curves}

Let us now consider the case when $P$ and $Q$ are curves. For convenience we will assume that they are $C_1$ continuous (i.e. continuous curves with well-defined tangent vectors everywhere). While it is possible to define the kernel distance between them by treating the curves as (infinite) sets of points, this approach ignores the fact that we would like to match not only the locations of points but the curve gradients. 

Let $t_P(p)$ denote a unit tangent vector at the point $p \in P$ on the curve. We can then define the pointwise similarity between the two points $p \in P$ and $q \in Q$ as the function $k(p,q) = K(p,q)\langle t_P(p), t_Q(q) \rangle$. The cross-similarity between $P$ and $Q$ can then be defined as 
\begin{equation}
\kappa(P, Q) = \int_P \int_Q K(p, q) \langle t_P(p), t_Q(q) \rangle.
\label{eq:curves}
\end{equation}

\subsection{Surfaces}
The same construction as above can be applied to orientable surfaces of co-dimension 1 in $\reals^d$, with the difference being that we use a unit \emph{normal} at a point (which is a vector), in place of its tangent.

\section{Reproducing Kernel Hilbert Spaces}
\label{sec:rkhs}

Not all similarity functions $K$ yield a distance metric. Consider the \emph{box similarity} given by 
\[ 
K(x,y) = \begin{cases} 1 & \|x - y\| \le 2 \\ 0 & \text{otherwise.}\end{cases} 
\]
Let $A$ be a set of three points all lying close to the origin in $\reals^1$. Fix $B$ to be the set $\lbrace (-1-\epsilon), (1+\epsilon), (0)\rbrace$. It is easy to verify that $D_K^2(A,B) = -2$, and thus $D_K(A,B)$ is not even defined. 

There is a simple sufficient condition on $K$ to ensure that $D_K$ to be a metric:
\begin{center}
  \emph{$K$ must be a positive definite kernel.}
\end{center}
For those well-versed in the theory of reproducing kernel Hilbert spaces~\cite{Aronszajn1950}, this observation follows easily by replacing the $K(x,y)$ terms by the appropriate inner products in the feature space. For those unfamiliar with this theory, an explanation follows. The material here is cobbled together from Daum\'{e}'s\cite{hal} and Jordan's\cite{jordan-lecnotes} notes on RKHSs. 

\subsection{Positive Definite Matrices}

Positive definite kernels generalizes the idea of a positive definite matrix. Let us start with a simple matrix kernel: $K(x,y) = x^\top A y$. If we set the matrix $A$ to be identity, we get the standard Euclidean inner product $\langle x, y\rangle$. In general however, we can set $A$ to be any positive definite matrix and guarantee that the resulting function $x^\top A y$ satisfies the properties of an inner product. 

\paragraph{A lifting map.}
Any positive definite matrix $A$ can be decomposed as $A = B^\top B$. Substituting this back into the bilinear form $x^T A y$, we get the expression $x^T B^T B y$, which we can rewrite as $\langle B x, B y\rangle$. If we now define the linear map $\Phi(x) = Bx$, we can rewrite the original matrix kernel as 
\[ K(x,y) = \langle \Phi(x), \Phi(y) \rangle. \]
There is further structure in the mapping $\Phi(x)$ that we can explore. Since $K$ is symmetric as well as real-valued, it can be written in the form $K = Q \Lambda Q^\top$ where the columns of $Q$ are eigenvectors of $K$, and $\Lambda$ is a diagonal matrix consisting of the eigenvalues of $K$. Since $K$ is positive definite, all the elements of $\Lambda$ are positive, and we can thus set $B = \Lambda^{1/2} Q^\top$, which means that each coordinate of $\Phi(x)$ is of the form $\sqrt{\lambda} \langle v, x \rangle$, where $v$ is an eigenvector of $K$ and $\lambda$ is the associated eigenvalue. More generally, this means that the eigenvectors of $K$ form an orthonormal basis describing the vector space $\Phi(x)$. 

\subsection{General Positive Definite Kernels}
If $K$ is a general function (i.e not a matrix), much of the above theory carries over in a similar manner, where instead of vector spaces we now reason about function spaces. 

\bignote{\small
\begin{center}
\textrm{A note on function spaces.}                                 
\end{center}

While not entirely rigorous, it is helpful to think of function spaces as merely infinite-dimensional vector spaces. A function $f : X \rightarrow \reals$ can be thought of as a vector whose ``dimensions'' are indexed by elements of $X$, so that $f(x) = c$ can be translated as ``the $x$th coordinate of the vector $f$ is $c$.'' Similarly, the function $K : X \times X \rightarrow \reals$ can be interpreted as a matrix indexed by elements of $X$ where the $(x,y)$th entry is $K(x,y)$. 

Integration can then be viewed as a product. For example, the integral $\int K(x,y) f(y) dy$ can be interpreted as a generalized matrix product between the ``matrix'' $K$ and the ``vector'' $f$. We will also look at ``eigenfunctions'' as the generalization of eigenvectors. 

}

We first define a positive definite kernel. 

\begin{defn}
A function $K : X \times X \rightarrow \reals$ is a positive definite kernel if for any $n$ and any set $\{x_1, x_2, \ldots, x_n\} \subset X$, the matrix $A = (a_{ij} = K(x_i, x_j))$ is positive definite. 
\end{defn}

Any positive definite kernel induces a Hilbert space (the \emph{reproducing kernel Hilbert space}) $\f{H}$. There are two different constructions yielding $\f{H}$. We will focus here on the method based on Mercer's theorem, since that yields an orthonormal representation of $\f{H}$ that will be useful.

A matrix $K$ defines the linear operator $T_K(f) = Kf$, here operating on a vector $f$ and producing another vector.  
Similarly, a positive definite kernel $K$ defines a linear operator on functions given by   
\[ 
[T_K(f)](\cdot) = \int K(\cdot, y) f(y) dy.
\]
Again in this continuous setting, $T_K$ is an operator that now takes a function $f$ as input and returns a function $[T_K(f)](\cdot)$.
The operator $T_K$ is linear by virtue of the linearity of the integral, and so it has eigenvalues and eigenfunctions. 

Mercer's theorem can then be seen as the analog of the above decomposition of the positive definite \emph{matrix} $A$ as $B^\top B, B = \Lambda^{1/2} Q^\top$. 
\begin{theorem}[Mercer]
  If $K$ is a continuous symmetric positive definite kernel, then there is an orthonormal basis $\lbrace v_i \rbrace$ consisting of eigenfunctions of $T_K$ (with associated eigenvalues $\lambda_i$) such that 
  \[ K(x,y) = \sum_0^\infty \lambda_i v_i(x)v_i(y). \]
\end{theorem}
The trick here is in thinking of $v_i(x)$ as the $x$th ``coordinate'' of the ``eigenvector'' $v_i$. 

The main consequence of this result is that we can now describe $\f{H}$ as the vector space spanned by the $v_i$, with the associated inner product given by 
\[ 
\bigl\langle \sum c_i v_i, \sum d_j v_j \bigr\rangle_\f{H} = \sum \frac{c_i d_i}{\lambda_i}. 
\]
By scaling the coordinates $c_i$ by $\sqrt{\lambda_i}$, we can use the Euclidean inner product instead, giving us a Euclidean space.
We summarize with the following statement, which captures the key properties of positive definite kernels. 

\begin{theorem}
  For any positive definite kernel $K : X \times X \rightarrow \reals$, there exists a Euclidean space $\f{H}$ and a \emph{lifting map} $\Phi : X \rightarrow \f{H}$ such that 
  \[ K(x, y) = \langle \Phi(x), \Phi(y) \rangle. \]
\end{theorem}

\subsection{The Kernel Distance as a Hilbertian metric}

We can now rewrite the kernel distance $D_K$ in terms of the associated lifting map $\Phi$. Using the linearity of the inner product, it is easy to show that 
\[ D_K(P, Q) = \Bigl\| \sum_{p \in P} \Phi(p) - \sum_{q \in Q} \Phi(q) \Bigr\|. \]
This of course immediately implies that $D_K$ is a pseudometric. A further technical condition on $K$ is needed to ensure that $D_K(P, Q) = 0 \Rightarrow P = Q$; the reader is referred to~\cite{hilbert} for more details. 

There are four important consequences of this recasting of the kernel distance. 
\begin{enumerate}
  \item The kernel distance embeds \emph{isometrically} in a Euclidean space. While in general $\f{H}$ might be infinite dimensional, the Hilbert space structure implies that for any finite collection of inputs, the effective dimensionality of the space can be reduced via projections to a much smaller finite number~\cite{current,Rahimi2007,Yang2003}.
  \item Most analysis problems are ``easier'' in Euclidean spaces. This includes problems like near-neighbor finding and clustering. The embedding of the kernel distance in such a space means that we now have access to a number of tools for analyzing collections of shapes.  Indeed, a complicated shape can be represented as a single vector $\bar\Phi(P) = \sum_{p \in P} \Phi(p)$ in the RKHS~\cite{praman}.
  \item The embedding ``linearizes'' the metric by mapping the input space to a vector space. This is indeed the primary reason for the popularity of kernel methods in machine learning. In the context of shape analysis, it means that many problems in the analysis of shape (finding consensus, averages, etc) can be solved easily by exploiting the linear structure of the lifted space.
  \item The complexity of computing the kernel distance is reduced significantly. If we assume that $\f{H}$ is approximated to within the desired error by a space of fixed dimension $\rho$, then in this space, computing the kernel distance between two point sets of total size $n$ takes $O(n\rho)$ time, instead of $\Theta(n^2)$ time. Since $\rho$ will in general be logarithmic in $n$ (or even independent of $n$), this is a significant improvement.
\end{enumerate}

\section{Mathematical Motivation}
\label{sec:motivation}

There are two distinct motivations that yield the kernel distance as formulated in \eqref{eq:current}. The first motivation is based on the desire to \emph{metrize} distributions, to construct a metric on distributions such that a convergent sequence with respect to this distance metric also converges in distribution. The second motivation comes from shape analysis in an attempt to use a geometric measure-theoretic view of shapes to define a distance between them that does not rely on explicit correspondences between features of shapes. 

Both of these approaches rely on the idea of ``test functions'' and ``dual spaces'', which we explore next. 

\subsection{Dual Vector Spaces And Test Functions}
Let $V$ be a vector space, and let $f$ be a continuous linear functional over $V$. In other words, $f$ is a linear function that takes an element $v \in V$, and returns a scalar $f(v)$. The space of all such $f$ is itself a vector space, and is called the \emph{dual} of $V$, denoted $V^*$.  
A very simple discrete example has $V$ and $V^*$ represent the columns and rows, respectively, of a fixed matrix.  
We can use structures defined on $V^*$ to construct similar structures on $V$; this is particularly useful when $V$ is an ill-formed space, but $V^*$ has more structure. 

Consider another simple example. Let $V$ be $\reals^d$, in which case the dual space $V^*$ is also $\reals^d$. Suppose we wish to define a norm on $V$. We can do this via \emph{pullback} from a norm on $V^*$ by the following construction:
\[ 
\|v \| = \sup_{ w \in V^*, \|w\| \le 1} |w(v)|. 
\]
It is not hard to verify that this indeed satisfies the properties of a norm. In fact, if we set the norm on $V^*$ to be an $\ell_p$ norm with $p \ge 1$, then the resulting norm $\| \cdot \|$ on $V$ is $\ell_q$, where $1/p + 1/q = 1$ (the dual norm). 

Notice that choosing different subsets of $V^*$ (by changing the norm) allows us to generate different norms on $V$. The case of $\ell_2$ is particularly instructive. Here, the set of all $w$ such that $\|w\|\le 1$ is the set of all vectors in the unit ball, and since $|w(v)|$ is increased by making $w$ larger, we can assume without loss of generality that $\|w\| = 1$, which means that we are considering all directions in $\reals^d$. In this case, $w(v)$ is merely $\langle w,v \rangle$, the projection of $v$ onto the direction $w$, and so the maximization over $w$ returns precisely the $\ell_2$ norm of $v$.  

Norms yield metrics in the usual way, by setting $d(x,y) = \|x-y\|$. Continuing the same construction, we can define a metric on $V$ by the construction
\[ 
d(v, v') = \sup_{w \in V^*, \|w\|\le 1} |w(v) - w(v')|. 
\]
Again returning to our example in $\ell_2$, this has a nice geometric interpretation. Since each $w$ is a direction in $\reals^d$, the expression being maximized is merely the length of the vector $\overline{vv'}$ when projected onto $w$. Clearly, maximizing this yields the actual $\ell_2$ distance between $v$ and $v'$.

\subsection{Metrizing Measures}
\label{sec:metrize}
With the above construction in mind, we can now build a metric over probability measures. The general construction, deemed an \emph{integral probability measure} by Muller~\cite{muller}, takes two probability measures $P$ and $Q$ over a space $M$, and defines the distance%
\footnote{Note that $\int f dP$ is a sparser notation for the more explicit representation of the intregral $\int_x f(x) P(x) dx$ or $\int f(x) dP(x)$.} 
\[ 
d_\f{F}(P,Q) = \sup_{f \in \f{F}} \left|\int f dP - \int f dQ\right|.
\]
Here, the class $\f{F}$ is the dual space to the space of probability measures, and consists of a class of real-valued bounded measurable functions over $M$. We can think of $\int f dP$ as the action $f(P)$ of $f$ on $P$; this is linear since the integral operator is linear. 

Many interesting distances between distributions can be generated by picking a particular subset of $\f{F}$. Some well-known cases:
\begin{itemize}
  \item If $\f{F} = \lbrace f \mid \|f\|_\infty \le 1\rbrace$, then $d_\f{F}(P,Q)$ is the $\ell_1$ distance between $P$ and $Q$. 
  \item If $M$, the domain of $f$, is endowed with a metric $\rho$, and we define $\|f\|_L = \sup \lbrace \frac{f(x) - f(y)}{\rho(x,y)}, x \ne y\rbrace$, then the resulting distance is the Kantorovich metric, also known as the Earth mover's distance~\cite{emd}.
\end{itemize}

Finally, and this is the case of relevance here, we can set $\f{F} = \lbrace f \mid \|f\|_\f{H} \le 1\rbrace$, where $\f{H}$ is a reproducing kernel Hilbert space. In this setting, it can be shown that the resulting distance can be rewritten as 
\[ 
D_K(P, Q) = \left\| \int K(\cdot, x) dP(x) - \int K(\cdot, x) dQ(x)\right\|_\f{H} 
\]
which, after some expansion and simplification, yields the familiar form 
\[ 
D^2_K(P, Q) = \int \int K(x,y) dP(x) dP(y) + \int \int K(x,y) dQ(x) dQ(y) - 2 \int \int K(x,y) dP(x) dQ(y) 
\]
which brings us back to \eqref{eq:current}. 

Note that by choosing the measures $P$ and $Q$ appropriately, we can generate the different variants of the kernel distance described in Section~\ref{sec:variations}. 

\subsection{Currents: Distributions with geometry}
\label{sec:current-dual}

We now move to the problem of comparing shapes, specifically point sets, curves and surfaces. The key viewpoint we will take is to treat the shape $S$ as a probability measure.  For point sets, this approach is easy to see, and for curves and surfaces, an appropriate discretization yields the desired probability measure. 
Once this is done, we can invoke the techniques from Section~\ref{sec:metrize} to define a distance metric between two shapes. 

For point sets, this construction is sufficient; however, curves and surfaces have more information than merely the spatial location of points. The well known example comparing the Hausdorff distance to the Frech\'{e}t distance for comparing curves illustrates that a proper comparison of curves requires not only spatial information, but also information about \emph{orientation}, and this continues to be true for surfaces. 

Geometric measure theory is the discipline that develops the tools to do this, and develops the idea of a \emph{current} as the geometric measure-theoretic analog of a surface that elegantly encodes the orientation information of the surface.  The idea of normal vectors generalizes to higher dimensions with $k$-forms.


\bignote{
\small
  \begin{center}
    \emph{Wedge Products, $k$-vectors, and $k$-forms}
  \end{center}

The \emph{cross product} of two vectors in  $\reals^3$ is a well-defined function whose main characteristics are that $v \times v = 0$ and $v \times w + w \times v = 0$. The \emph{wedge product} (denoted by $\wedge$) generalizes the cross product to arbitrary vectors $v \in \reals^d$. It has the property that $v \wedge v = 0$ (which implies that $v \wedge w = - w \wedge v$  \footnote{This can be shown by expanding $(w + v) \wedge (w + v) = 0$.}). A simple \emph{$k$-vector} is defined as the wedge product of $k$ vectors: $w = v_1 \wedge v_2 \cdots \wedge v_k$, and the space of all linear combinations of $k$-vectors is denoted $\wedge_k(\reals^d)$. Note that $\wedge_1(\reals^d) = \reals^d$. $k$-vectors generalize the idea of a tangent vector, by capturing in one expression a $k$-dimensional subspace of the tangent space at a point. 

A differential form can now be understand as a linear functional over the $k$-vectors. Formally, a \emph{$k$-form} $\omega$ is an assignment of a linear functional $\omega_x$ at each point $x$ on the manifold. This linear functional $\omega_x$ is a linear mapping from $\wedge_k(\reals^d)$ to $\reals$, and the space of all such functionals is denoted $\wedge^k(\reals^d)$. Since $k$-forms and $k$-vectors are dual by construction, we will use the expression $\omega(v)$ or $\langle w, v\rangle$ to denote the induced action of one on the other. 
}


We can integrate a $k$-form $\omega$ over a $k$-dimensional oriented manifold $S$.  Using $dS$ to denote the appropriate ``tangent space displacement'' (a generalized tangent vector) on the surface, we can write this integral as $\int \omega(dS)$ or $\int \omega dS$ when the action is understood. By integrating on the $k$-vector $dS$ at each point $p \in S$, instead of just the point $p$, we capture the orientation information of the manifold $S$, not just the spatial information. 
By duality, this can be seen as an action of $S$ on $\omega$, and leads us (finally) to the definition of a current:

\begin{defn}
  A $k$-current is a continuous linear functional over the space of $k$-forms. 
\end{defn}

Currents generalize oriented manifolds. They allow us to account for irregularities and sharp edges in a manifold, at the cost of including elements that may not look like standard manifolds. Very crudely, the relation between currents and oriented manifolds is akin to the relationship between (Schwarz) distributions and probability measures. 

We can now define a norm on the space of currents much as before, by placing a norm on the space of $k$-forms and pulling it back. The equivalent of the total variation norm (by computing $\sup \omega(v), \|\omega\|_\infty \le 1$) is called the \emph{mass norm}; unfortunately, much like the total variation, it does not capture variations between currents in a useful manner. For example, given two curves with disjoint support, the distance induced by the mass norm is $2$, irrespective of the actual shape of the curves. This is similar to how the total variation distance between two distributions is always $2$ if they have disjoint support.

By setting the space of test functions to be the unit ball in an RKHS as before, we retrieve the kernel distance as described in Section~\ref{sec:variations}. 
For instance, consider two curves $S$ and $T$ so the $k$-vectors $dS$ and $dT$ are the tangent vectors $t_S(\cdot)$ and $t_T(\cdot)$.  
We can expand
\[
D_K(S,T) = \left \| \int K(\cdot, x) dS(x) - \int K(\cdot, x) dT(x) \right \|_\f{H}
\]
as
\begin{align*}
D^2_K(S, T) 
&= 
\int \int  (dS(x))^T K(x,y) dS(y) + \int \int (dT(x))^T K(x,y) dT(y) - 2 \int \int  (dS(x))^T K(x,y) dT(y) 
\\ &=
\int_S \int_S  (t_S(x))^T K(x,y) t_S(y) + \int_T \int_T (t_T(x))^T K(x,y) t_T(y) - 2 \int_S \int_T  (t_S(x))^T K(x,y) t_T(y), 
\end{align*}
where $(\cdot)^T$ represent a vector transpose.  Since $K(x,y)$ is a scalar, we can factor it out of each term
\[
D^2_K(S, T) 
= 
\int_S \int_S  K(x,y) \langle t_S(x), t_S(y) \rangle + \int_T \int_T K(x,y) \langle t_T(x), t_T(y) \rangle - 2 \int_S \int_T  K(x,y) \langle t_S(x), t_T(y) \rangle,
\]
precisely as in the similarity term from (\ref{eq:curves}).  This derivation of $D_K$ has led to the alternate name current distance.  

\bibliographystyle{acm}
\bibliography{refs}
\end{document}